\documentstyle[preprint,12pt,aps,eqsecnum,prbbib]{revtex}

\begin{document}

\title{Electrically neutral Dirac particles in the presence of\\
        external fields: exact solutions\\}
\author{German V. Shishkin}
\address{Department of Theoretical Physics, Byelorussian State
University\\
Minsk 220050, Byelorussia\\}

\author{V\'{\i}ctor M. Villalba}
\address{Centro de F\'{\i}sica\\
Instituto Venezolano de Investigaciones Cient\'{\i}ficas, IVIC\\
Apdo 21827, Caracas 1020-A, Venezuela}

\maketitle
\begin{abstract}
In the present article we present exact solutions of the Dirac equation
for electric
neutral particles with anomalous electric and magnetic moments. Using
the algebraic
method of separation of variables, the Dirac equation is separated in
cartesian, cylindrical
and spherical coordinates, and exact solutions are obtained in terms of
special functions.
\end{abstract}
\pacs{11.10Qr}

\section{Introduction}
\label{sec:level1}

The experimental data on the existence of anomalous magnetic moment and
electric dipole moment for the Dirac ~particle\cite{Particle} require
for their complete understanding an exact description of the
corresponding
single-particle states, ~i.e., exact solutions of the Dirac equation
with
non-minimal interactions. In some physical situations the interaction
between the anomalous magnetic moment and electric dipole moments with
the electromagnetic tensor can be solved exactly.

In fact, if for the electron and the muon (and apparently also for the
$\tau
$-lepton) the deviation of the magnetic moment from the Bohr's magneton
and
the value of the ~electric dipole moment are extremely small, on the
other hand, for the neutrino ($\nu _e,\nu _\mu ,\nu _\tau $) despite the
magnetic moment is very small it is one of their ~nonzero characteristic
parameters. Also, we have that for the hadrons the values of the normal
and
anomalous magnetic moments are of the same order of magnitude. For all
the
hadrons with large mean life it has been established, when experiments
make
it possible, the presence of electric dipole moment. For electrically
neutral hadrons (neutron, $\Lambda ,\Sigma ^0,\Xi ^0$ hyperons) the
interaction between the anomalous magnetic moment with the external
field
becomes significant.

Among the above mentioned particles, the neutron plays a privileged
position
because of their physical applications. In fact, one of the most crucial
problems of the present century is the security of the nuclear energy.
One
of the most important ways for analyzing the kinematic parameters of a
nuclear reactor is the injection of neutrons and the subsequent
measurement
of the decay on time of the net population inside the reactor. The
mathematical models for this problem are still primitive: the momentum
is
replaced by a time dependent delta function, and in general, the space
distribution of neutrons in impulse is unknown for the experimenter.
Since
neutron beams are ~controlled by means of magnetic and electric fields,
in
order to understand the space distribution of neutrons it would be of
help
to know the corresponding wave equations, i.e, the solutions of the
Dirac
equation for the neutron.

In view of all the above mentioned, we will examine in the present
article
exact solutions of the Dirac equation for electrically neutral particles
with non minimal interaction with an external electromagnetic field,
that
is, we analyze the interaction between anomalous magnetic and electric
dipole moments with the ~electromagnetic tensor.

The amount of articles devoted to the study of this problem is scarce,
the
reason is that, for solving the Dirac equation, as general rule, a
complete
separation of variables is required and, the inclusion of tensor field
functions in the Dirac equation dramatically restricts the
{}~possibilities of
separation of variables \cite{Shishkin1,Shishkin2,Shishkin3,Shishkin4}
In
this case, the ~possibilities are even more limited that those obtained
for
the Dirac equation minimally coupled to the electromagnetic field\cite
{Shishkin5}, or in the presence of gravitational fields \cite{Shishkin6}
This one is because the gravitational field goes into the ~Dirac
equation in
a geometric way, via the Lame's coefficients, and no additional matrices
in
the equation if we choose to work in a diagonal tetrad gauge. Regarding
the
vector electromagnetic fields, the inclusion of ~them in the Dirac
equation
via the generalized momentum (minimal coupling), does not conduce to the
apparition of new matrices, and \cite{Shishkin5} the presence of tensor
fields terms of the form $g\gamma ^m\gamma ^nT_{mn}$ introduces in the
Dirac
equation a new functional dependence. The difficulties that arise are
those
we find when we include non geometrized fields like scalar,
pseudo-scalar or
pseudo-vector fields. \cite{Shishkin3,Shishkin4,Cook}

Now, we proceed to mention the most relevant exact solutions of the
Dirac
equation with anomalous moment reported in the literature, among them we
have the constant magnetic field problem solved by
Strocchi\cite{Strocchi}
and by Ternov{\it \ et al }\cite{Ternov}, a generalization of the
Volkov's
problem was found by Ternov {\it et al }\cite{Ternov2,Ternov3}, and more
recently this problem has been revisited by Barut \cite{Barut1}. The
problem
in spherical coordinates for the central field has been studied by
Ternov
\cite{Ternov4} and Barut \cite{Barut2,Barut3}, a good review on exact
solutions of the Dirac equation with anomalous interaction can be found
in
\cite{BagrovL}.

In the present article we will discuss this problem from a unified point
of
view based on the algebraic method of separation of variables \cite
{Shishkin5,Shishkin6}, which consists in reducing the original Dirac
equation to a sum of two commuting differential operators as follows

\begin{equation}
\label{sep}\left\{ {\bf \hat K}_1+{\bf \hat K}_2\right\} \Phi =0,\quad
\left[ {\bf \hat K}_1{\bf \hat K}_2\right] _{-}=0
\end{equation}
where each operator depends on a different set of variables. Then, we
reduce
the problem of solving the original equation to obtain solutions for
${\bf %
\hat K}_1\Phi =\lambda \Phi =-{\bf \hat K}_2\Phi $ where $\lambda $ is a
constant of separation. Now, applying the scheme (\ref{sep}) for ${\bf
\hat K%
}_1\Phi =\lambda \Phi $ and ${\bf \hat K}_2\Phi +\lambda \Phi =0$ and
iterating this process, we are able to reduce a system coupled partial
differential equations to four systems of ordinary differential
equations.
Regretfully, not always a complete separation is possible because of the
structure of the metric functions and the form of the external fields.

The article is structured as follows, in Sec II, the Dirac equation with
anomalous electric and magnetic moments is separated and some exact
solutions are exhibited. In Sec. III, the Dirac equation is separated in
cylindrical and spherical coordinates, also, some exact solutions in
terms
of special functions are presented. In Sec. IV, we present a particular
configuration where Dirac equation is separable and exact solutions are
obtained. Finally, in the conclusions, we summarize and discuss our
results.

\section{Cartesian coordinates}

In this section we proceed to separate variables and to find exact
solutions
of the Dirac equation with anomalous electric and magnetic moments when
this
interaction can be written in Cartesian coordinates. Then, for a neutral
particle interacting with a magnetic field, Dirac equation in Cartesian
coordinates reads

\begin{equation}
\label{Dirac}\left\{ \gamma ^0\partial _t+\gamma ^1\partial _x+\gamma
^2\partial _y+\gamma ^3\partial _z+m+g\epsilon _{ijk}\gamma ^i\gamma
^jH_k\right\} \Psi =0
\end{equation}

Applying the algebraic method of separation of
variables\cite{Shishkin5,Shishkin6},
we can write Eq. (%
\ref{Dirac}) as follows

\begin{equation}
\left( \hat K_{xy}+\hat K_{zt}\right) \Phi =0,\quad \hat K_{xy}\Phi
=\lambda
\Phi =-\hat K_{zt}\Phi
\end{equation}
with,
\begin{equation}
\label{operator}\hat K_{xy}=\gamma ^2\partial _x-\gamma ^1\partial
_y+\gamma
^1\gamma ^2m\ -gH_z(x,y),
\end{equation}
\begin{equation}
\label{operator2}\hat K_{zt}=\gamma ^1\gamma ^2\gamma ^3\partial
_z+\gamma
^1\gamma ^2\gamma ^0\partial _0-g\tilde H(z,t),
\end{equation}
and
\begin{equation}
\Phi =\gamma ^1\gamma ^2\Psi
\end{equation}
where the magnetic vector $H_z$ can be written as, \

\begin{equation}
\label{H}H_z=H_z(x,y)+\tilde H_z(z,t)
\end{equation}
applying the condition $\nabla .H=0,$ on ($\ref{H})$ we find that
$\tilde
H_z=\tilde H_k(t)$, and if we fix our attention to time independent
fields
this term can be omitted. In this way, $\lambda \Phi =-\hat K_{zt}\Phi $
becomes in an algebraic equation that establishes the relationship
between
the different spinor components of $\Phi $ and $\Psi ,$ and the value of
the
constant of separation $\lambda $ , which satisfies the relation
$\lambda
^2=k_z^2-E^2.$

Choosing to work in the following representation for the gamma matrices

\begin{equation}
\gamma ^2=\pmatrix{1&0\cr0&-1},\gamma ^0=%
\pmatrix{0& i\sigma^{2}\cr i\sigma^{2}& 0\cr},\gamma ^1=%
\pmatrix{0& \sigma^{1}\cr \sigma^{1}& 0\cr},\gamma ^3=%
\pmatrix{0& \sigma^{3}\cr \sigma^{3}& 0\cr},
\end{equation}
we have that,
\begin{equation}
\left\{ \gamma ^2\partial _1-\gamma ^1\partial _2+m\gamma ^1\gamma
^2-gH_z(x)-\lambda \right\} \Phi ~=~0
\end{equation}
takes the form
\begin{equation}
\label{ord1}(\partial _x-gH_z(x)-\lambda )\Phi _{1,2}-(m+ik_y)\Phi
_{4,3}~=~0
\end{equation}
\begin{equation}
\label{ord2}(\partial _x+gH_z(x)+\lambda )\Phi _{4,3}-(m-ik_y)\Phi
_{1,2}~=~0
\end{equation}
on the other hand, from $\lambda \Phi =-\hat K_{zt}\Phi $ we find,

\begin{equation}
\Phi ~=~\pmatrix{\Phi_{1}\cr \Phi_{2}\cr \Phi_{3}\cr \Phi_{4}\cr}~=~%
\pmatrix{\Phi_{1}\cr (i\lambda -E)/k_{z}\Phi_{1}\cr (i\lambda
-E)/k_{z}\Phi_{4}\cr \Phi_{4}}
\end{equation}

\noindent the coupled system of equations (\ref{ord1}) and (\ref{ord2})
can
be solved in terms of special functions for some values of the magnetic
field $H_z(x).$ Among them we have,
\begin{equation}
(a)\ H_z=\beta ,\ (b)\ H_z=\ \beta x,\ \ (c)\ H_z~=~\beta /x,\quad \quad
(d)\ H_z=\ \beta \exp (\eta x),
\end{equation}
first, let us consider the simplest case of a constant magnetic field
$(a)\
H_z=\beta $. Then, the solution of the system reads:
\begin{equation}
\Phi _1=c_1\exp (\sqrt{m^2+k_y^2+(g\beta +\lambda )^2}x)+c_2\exp
(-\sqrt{%
m^2+k_y^2+(g\beta +\lambda )^2}x)
\end{equation}

\begin{eqnarray}
\Phi_{2}=c_{1}\frac{\sqrt{m^2+k^{2}_{y}+(gH-\lambda)^2}-gH+\lambda}{m+ik_{y}}\exp(\sqrt{m^2+k^{2}_{y}+(gH-\lambda)^2}x)\\
  \nonumber
-c_{2}\frac{\sqrt{m^2+k^{2}_{y}+(gH-\lambda)^2}+gH+\lambda}{m+ik_{y}}\exp(\sqrt{m^2+k^{2}_{y}+(gH-\lambda)^2}x)
\end{eqnarray}

$(b)\ H_z=\beta x$ For this linear magnetic field we have that the
system of
equations (\ref{ord1})(\ref{ord2}) can be reduced to the following two
second order differential equations
\begin{equation}
\label{lin1}[(\partial _x+g\beta x+\lambda )(\partial _x-g\beta
x-\lambda
)-(m^2+k_y^2)]\Phi _1=0
\end{equation}
\begin{equation}
\label{lin2}[(\partial _x-g\beta x-\lambda )(\partial _x+g\beta
x+\lambda
)-(m^2+k_y^2)]\Phi _4=0
\end{equation}
after making the change of variables
\begin{equation}
g\beta x+\lambda =\left( {\frac{g\beta }2}\right) ^{1/2}y
\end{equation}
we have that the equations (\ref{lin1}) and (\ref{lin2}) take the form:
\begin{equation}
\label{s1}\{\frac{d^2}{dy^2}-\frac{y^2}4-\frac{g\beta
+m^2+k_y^2}{2g\beta }%
\}\Phi _1=0
\end{equation}
\begin{equation}
\label{s2}\{\frac{d^2}{dy^2}-\frac{y^2}4-\frac{-g\beta
+m^2+k_y^2}{2g\beta }%
\}\Phi _4=0
\end{equation}
the solution of the parabolic cylinder equation (\ref{s1}) can be
written in
terms of confluent hypergeometric functions $M(a,b,z)$ \cite{Abramowitz}

\begin{equation}
\Phi _1=c_1\exp (-y^2/4)M(\frac 12+\frac{m^2+k_y^2}{4g\beta },\ \frac
12,\
\frac{y^2}2)+c_2y\exp (-y^2/4)M(1+\frac{m^2+k_y^2}{4g\beta },\ \frac
32,\
\frac{y^2}2)
\end{equation}
using (\ref{ord1}) and the recurrence relation
\begin{equation}
(b-1)M(a-1,b-1,z)=(b-1-z)M(a,b,z)+zM^{\prime }(a,b,z)
\end{equation}
we obtain
\begin{equation}
\delta \Phi _4=c_1y\exp (-y^2/4)M(\frac 12+\frac{m^2+k_y^2}{4g\beta },\
\frac 32,\ \frac{y^2}2)+c_2\exp (-y^2/4)M(\frac{m^2+k_y^2}{4g\beta },\
\frac
12,\ \frac{y^2}2)
\end{equation}
where the constant $\delta $ reads,
\begin{equation}
\delta =\frac{m+ik_y}{\sqrt{2}(g\beta )^{1/2}}
\end{equation}
The third case corresponds to the magnetic field $(c)\ H_z~=~\beta /x.$
Substituting this expression into (\ref{ord1}) and (\ref{ord2}) we
obtain
\begin{equation}
\label{firsto}\left\{ \frac{d^2}{dx^2}\ +\ \frac{2g\beta \lambda }x\
-\frac{%
g\beta (g\beta +1)}{x^2}\ -\ k_y^2\ -\lambda ^2-m^2\right\} \Phi _4=0
\end{equation}
\begin{equation}
\label{sec}\left\{ \frac{d^2}{dx^2}\ -\ \frac{2g\beta \lambda }x\
-\frac{%
g\beta (g\beta +1)}{x^2}\ -\ k_y^2\ -\lambda ^2-m^2\right\} \Phi _1=0
\end{equation}
the solutions of (\ref{firsto}) and (\ref{sec}) also can be expressed in
terms of confluent hypergeometric functions as follows,
\begin{equation}
\Phi _4=\exp (-\frac{y^2}2)\left\{ a_0y^{g\beta +1}M(1+g\beta -\tilde
k,\
2+2g\beta ,\ y)\ +a_1\ y^{-g\beta }M(-g\beta -\tilde k,-2g\beta ,\
y)\right\}
\end{equation}
\begin{equation}
\Phi _1=\exp (-\frac{y^2}2)\left\{ b_0y^{g\beta }M(g\beta -\tilde k,\
2g\beta ,\ y)\ +b_1\ y^{1-g\beta }M(1-g\beta -\tilde k,\ 2-2g\beta ,\
y)\right\}
\end{equation}
where
\begin{equation}
y=2\sqrt{\lambda ^2+k_y^2+m^2}x,\quad \tilde k=-\frac{g\beta \lambda
}{\sqrt{%
\lambda ^2+k_y^2+m^2}}
\end{equation}
and the coefficients $a_0,\ a_{1,}\ b_0$ and $b_1\ $satisfy the
following
relations
\begin{equation}
\frac{a_1}{b_1}=\frac{2i}{k_y-im}(1-2g\beta )\sqrt{\lambda ^2+k_y^2+m^2}
\end{equation}
\begin{equation}
\frac{a_0}{b_0}=\frac{(ik_y-m)}{2(2g\beta +1)\sqrt{\lambda
^2+k_y^2+m^2}}
\end{equation}
finally, we have the exponential depending magnetic field given by $(d)\
H_z=\ \beta \exp (\eta x).$ Here we have that, after making the change
of
variables $\mu =\exp (\eta x)$, the system of equations (\ref{ord1})
(\ref
{ord2}) takes the form
\begin{equation}
\label{exp1}(ik_y+m)\Phi _4\ -\ (\eta \mu \partial _\mu \ -(\lambda \
+\beta
\mu )\Phi _1=0
\end{equation}
\begin{equation}
\label{exp2}(ik_y-m)\Phi _1\ +\ (\eta \mu \partial _\mu \ +(\lambda \
+\beta
\mu )\Phi _4=0
\end{equation}
the system of equations (\ref{exp1}) (\ref{exp2}) is similar to the one
obtained when we solve the Dirac equation for an electron in a magnetic
field like\cite{Shishkin5} (d). Then we have that,

$$
\Phi _4=\exp (-\frac{\beta \mu }\eta )a_0\mu ^{i\frac{k_x}\eta }M(\frac
\lambda \eta +\frac{ik_x}\eta ,\frac{2ik_x}\eta +1,\ \frac{2\beta \mu
}\eta
)
$$
\begin{equation}
+\exp (-\frac{\beta \mu }\eta )a_1\mu ^{-i\frac{k_x}\eta }M(\frac
\lambda
\eta -\frac{ik_x}\eta ,-\frac{2ik_x}\eta +1,\frac{2\beta \mu }\eta
\end{equation}
$$
\Phi _1=ia_0\frac{ik_x+\lambda }{k_y+im}\exp (-\frac{\beta \mu }\eta
)\mu ^{i
\frac{k_x}\eta }M(\frac \lambda \eta +\frac{ik_x}\eta
+1,\frac{2ik_x}\eta
+1,\ \frac{2\beta \mu }\eta )\
$$
\begin{equation}
-ia_1\frac{ik_x-\lambda }{k_y+im}\exp (-\frac{\beta \mu }\eta )\mu
^{-i\frac{%
k_x}\eta }M(\frac \lambda \eta -\frac{ik_x}\eta +1,-\frac{2ik_x}\eta +1,
\frac{2\beta \mu }\eta )
\end{equation}

\section{Exact solutions in cylindrical and spherical coordinates}

In this section we are going to find some exact solutions of the Dirac
equation with anomalous moment in cylindrical and spherical coordinates.
The Dirac equation in spherical coordinates for a chargeless particle
with
anomalous electric moment, expressed in the diagonal tetrad
gauge\cite{Shishkin5,Shishkin6}, reads,

\begin{equation}
\label{spher}\left\{ \gamma ^0\partial _0+\gamma ^1\partial _x+{\frac
\gamma
r}^2\partial _\vartheta +\frac{\gamma ^3}{r\sin \vartheta }+m+g\gamma
^1\gamma ^0E\right\} \Psi ~=0
\end{equation}
substituting into equation (\ref{spher}) the electric field given $E$ by
the
expression $E=\epsilon /r,$ and separating the angular variables from
the
radius and time, we have
\begin{equation}
\label{separated}\left[ \gamma ^0\partial _r\ +\gamma ^1\partial _t\
+m\gamma ^1\gamma ^0\ +\frac{g\epsilon }r\ +\frac kr\right] \Phi =0
\end{equation}
\begin{equation}
\label{Brill}\left[ \gamma ^2\partial _\vartheta \ +\frac{\gamma
^3}{\sin
\vartheta }\partial _\varphi \right] \gamma ^1\gamma ^0\Phi =k\Phi
\end{equation}
where $k$ is a constant of separation associated with the angular
dependence. It is worth noting that the operator (\ref{Brill})
correspond to
the one obtained in the separation of variables for the Dirac equation
in
the Schwarzschild metric\cite{Schr,Brill} when we work in the diagonal
tetrad gauge. The spinor $\Phi $ is related to $\Psi $ by means of the
expression, $\Phi =\gamma ^1\gamma ^0\Psi .$

\noindent Working in the standard representation of the gamma matrices,
\begin{equation}
\label{repr}\gamma ^0=\left(
\begin{array}{cc}
-i & 0 \\
0 & i
\end{array}
\right) ,\quad \gamma ^i=\left(
\begin{array}{cc}
0 & \sigma ^i \\
\sigma ^i & 0
\end{array}
\right) ,\ i=1,2,3
\end{equation}
and substituting (\ref{repr}) into (\ref{separated}) we obtain,

\begin{equation}
\label{B1}\left[ \frac d{dr}+\frac{i(g\epsilon +k)}r\right] \Phi
_1-(m-{\sf E%
})\Phi _4=0
\end{equation}

\begin{equation}
\label{B2}\left[ \frac d{dr}\ -\ \frac{i(g\epsilon +k)}r\right] \Phi _4\
+\
(m\ +{\sf E})\Phi _1=0
\end{equation}
where the constant ${\sf E\ }$is the eigenvalue of the operator
$i\partial _t
$ .After substituting  (\ref{B1}) into (\ref{B2}) we obtain a second
order
differential equation given by the expression
\begin{equation}
\left[ \frac{d^2}{dr^2}+\frac{(g\epsilon +\kappa )^2-i(g\epsilon +\kappa
)}{%
r^2}+m^2-E^2\right] \Phi _1=0
\end{equation}
whose solution\cite{Kamke} can be expressed in terms of $Z_\upsilon
(z)=\alpha J_\upsilon (z)+\beta N_\upsilon (z)$ where $J_\upsilon (z)$
and $%
N_\upsilon (z)$ are the Bessel and Neumann functions respectively, and
$%
\alpha $ and $\beta $ are arbitrary constants,

\begin{equation}
\label{B3}\Phi _1=a_0r^{\frac 12}Z_{\frac 12-i(g\epsilon
+k)}(\sqrt{m^2-{\sf %
E}^2}x)
\end{equation}
then, substituting (\ref{B3}) into (\ref{B1}) we obtain,
\begin{equation}
\Phi _4=\frac{\sqrt{m^2-{\sf E}^2}}{m-{\sf E}}a_0r^{\frac 12}Z_{-\frac
12-i(g\epsilon +k)}(\sqrt{m^2-{\sf E}^2}x)
\end{equation}
here, it is worth noting that, given the matrix structure of eq. (\ref
{separated}), the spinor components $\Phi _1$ and $\Phi _4$ are
proportional
to $\Phi _2$ and $\Phi _3$ respectively.

\noindent Now we are going to solve Eq. (\ref{spher}) when a constant
electric field $E=\epsilon $ along the radial direction is present. In
this
case, the set of separated equations reads:
\begin{equation}
\label{odin}\left[ \frac d{dr}+ig\epsilon +\frac{ik}r\right] \Phi
_1-(m-{\sf %
E})\Phi _4=0
\end{equation}
\begin{equation}
\label{dva}\left[ \frac d{dr}\ -ig\epsilon \ -\frac{ik}r\right] \Phi _4\
+\
(m\ +{\sf E})\Phi _1=0
\end{equation}
the solution of the coupled system of equations (\ref{odin})-(\ref{dva})
can
be written by the help of the Bessel functions,
\begin{equation}
\Phi _1=cr^{\frac 12}Z_{\frac 12-i\omega }(\sqrt{m^2-E^2}r)
\end{equation}
\begin{equation}
\frac{m-{\sf E}}{\sqrt{m^2-{\sf E}^2}}\Phi _4=cr^{\frac 12}Z_{-\frac
12-i\omega }(\sqrt{m^2-E^2}r)
\end{equation}

The Dirac equation in cylindrical coordinates expressed in the diagonal
tetrad gauge takes the form,

\begin{equation}
\label{Cyl}\left\{ \gamma ^0\partial _t+\gamma ^1\partial _r+{\frac
\gamma r}%
^2\partial _\vartheta +\gamma ^3\partial _z+m+g\gamma ^1\gamma
^2H\right\}
\Psi ~=0
\end{equation}
where we have included a term associated with the anomalous magnetic
moment. Eq. (\ref{Cyl}) can be written as a sum of two commuting
differential operators:
\begin{equation}
\label{prim}\left( \gamma ^2\partial _r-{\frac \gamma r}^1\partial
_\vartheta +m\gamma ^1\gamma ^2-gH+\lambda \right) \Phi ~=~0
\end{equation}
\begin{equation}
\label{spin}\left( \gamma ^1\gamma ^2\gamma ^3\partial _z\ +\gamma
^1\gamma
^2\gamma ^0\partial _0\right) \Phi =\lambda \Phi ,\quad \Psi =\gamma
^1\gamma ^2\Phi
\end{equation}
here, it is convenient to work in the following representation of the
Dirac
matrices,

\begin{equation}
\label{DM}\gamma ^1~=~\pmatrix{\sigma ^{1}& 0\cr 0& -\sigma^{1}\cr},\
\gamma
^2~=~\pmatrix{\sigma ^{2}& 0\cr 0& -\sigma^{2}\cr},\ \gamma ^3~=~%
\pmatrix{\sigma^{3}& 0\cr 0& -\sigma^{3}\cr},\ \gamma ^0~=~%
\pmatrix{0&i\cr i&0 \cr}
\end{equation}
then, after substituting (\ref{DM}) into (\ref{prim}) we arrive at,
\begin{equation}
\label{theta}(\sigma ^2\partial _r-{\frac \sigma r}^1\partial _\vartheta
+im\sigma ^3-gH+\lambda )\Theta ~=~0
\end{equation}
\begin{equation}
(-\sigma ^2\partial _r+{\frac \sigma r}^1\partial _\vartheta +im\sigma
^3-gH+\lambda )\chi ~=~0
\end{equation}
where, taking into account (\ref{spin}) we have
\begin{equation}
\Phi =\left(
\begin{array}{c}
\Theta  \\
\chi
\end{array}
\right) =\left(
\begin{array}{c}
\Theta  \\
\frac{iE\sigma ^3}{k_z-\lambda }\Theta
\end{array}
\right) ,\Theta =\left(
\begin{array}{c}
\Theta _1 \\
\Theta _2
\end{array}
\right)
\end{equation}
where $\lambda ^2=k_z^2-E^2$. Then, substituting the explicit
representation
of the Pauli matrices into (\ref{theta}) we obtain that the equations
governing the radial dependence of $\Theta $ are:
\begin{equation}
\label{u1}(d_r-{\frac{k_\vartheta }r})\Theta _1-(m+i(\lambda -gH))\Theta
_2=0
\end{equation}
\begin{equation}
\label{u2}(d_r+{\frac{k_\vartheta }r})\Theta _2-(m-i(\lambda -gH))\Theta
_1=0
\end{equation}
Here, there are two simple cases for which the system
(\ref{u1})-(\ref{u2})
can be solved exactly in terms of special functions. They are: a) the
constant magnetic field $H,$ and b) $H=\alpha /r.$

a) For this case we obtain the following second order differential
equations
\begin{equation}
\label{uno}\left( \frac{d^2}{dr^2}-\frac{k_\vartheta (k_\vartheta
+1)}{r^2}%
-m^2-(\lambda -gH)^2\right) \Phi _2=0
\end{equation}
\begin{equation}
\label{dos}\left( \frac{d^2}{dr^2}-\frac{k_\vartheta (k_\vartheta
-1)}{r^2}%
-m^2-(\lambda -gH)^2\right) \Phi _1=0
\end{equation}
whose solutions take the form\cite{Kamke}
\begin{equation}
\Phi _1=c_1r^{1/2}Z_{k_\vartheta -1/2}(i\sqrt{m^2+(\lambda -gH)^2}r)
\end{equation}
\begin{equation}
\Phi _2=c_2r^{1/2}Z_{k_\vartheta +1/2}(i\sqrt{m^2+(\lambda -gH)^2}r)
\end{equation}
where $Z_\alpha (z)$ is the general solution of the Bessel equation, and
the
constants$\ c_1$ and $c_2$ are related as follows,

\begin{equation}
c_1=ic_2\frac{m+i(\lambda -gH)}{\sqrt{m^2+(\lambda -gH)^2}}
\end{equation}
b) Substituting $H=\alpha /r$ into (\ref{u1})-(\ref{u2}) we get,

\begin{equation}
\label{h1}(d_r-{\frac{k_\vartheta }r})\Theta _1-(m+i\lambda
-i\frac{g\alpha }%
r)\Theta _2=0
\end{equation}
\begin{equation}
\label{h2}(d_r+{\frac{k_\vartheta }r})\Theta _2-(m-i\lambda
+i\frac{g\alpha }%
r)\Theta _1=0
\end{equation}
the system of equations (\ref{h1})-(\ref{h2}) resembles the one obtained
in
solving the Dirac equation with a Coulomb potential. Therefore it is
possible to express the solution of (\ref{h1})-(\ref{h2}) by means of
hypergeometric functions as follows:

\begin{equation}
\Theta _1=\sqrt{1-i\frac \lambda m}e^{-\frac \rho 2}(F_1(\rho )+F_2(\rho
))
\end{equation}
\begin{equation}
\Theta _1=\sqrt{1+i\frac \lambda m}e^{-\frac \rho 2}(F_1(\rho )-F_2(\rho
))
\end{equation}
where $\rho \ $$=2\sqrt{m^2+\lambda ^2}r,$ and $F_1$ and $F_2$ read:
\begin{equation}
F_2(\rho )=c_0\rho ^\gamma M(\gamma -\frac{g\alpha \lambda }{\sqrt{%
m^2+\lambda ^2}},\ 2\gamma +1,\ \rho )
\end{equation}
\begin{equation}
F_1(\rho )=-c_0\frac{\gamma -\frac{g\alpha \lambda }{\sqrt{m^2+\lambda
^2}}}{%
k_\vartheta -i\frac{g\alpha m}{\sqrt{m^2+\lambda ^2}}}\rho ^\gamma
M(\gamma
- \frac{g\alpha \lambda }{\sqrt{m^2+\lambda ^2}}+1,\ 2\gamma +1,\ \rho )
\end{equation}
where c$_0$ is a constant arbitrary . and $\gamma =\sqrt{k_\vartheta
^2+g^2\alpha ^2}$

Making use of the expression for the asymptotic behavior of the
confluent
hypergeometric functions for $\rho \rightarrow \infty $\cite{Abramowitz}

\begin{equation}
M(a,b,x)\rightarrow e^{-i\pi a}\frac{\Gamma (b)}{\Gamma (b-a)}\rho
^{-a}+
\frac{\Gamma (b)}{\Gamma (a)}\rho ^{a-b}e^\rho
\end{equation}
we have that the solutions of (\ref{h1})-(\ref{h2}) are regular al
infinity
only if
\begin{equation}
\frac 1{\Gamma (\gamma -\frac{g\alpha \lambda }{\sqrt{m^2+\lambda
^2}})}=0
\end{equation}
or equivalently,
\begin{equation}
\label{energia}E^2=k_z^2-\frac{m^2}{\frac{g^2\alpha ^2}{(\gamma
+n)^2}-1}
\end{equation}
where $n$ =0, 1, 2, ....., and $E$ is the energy.

the expression (\ref{energia}) takes a very simple form when $n=0$,
\begin{equation}
E^2=k_z^2+m^2+\left( \frac{g\alpha m}{k_\vartheta }\right) ^2
\end{equation}

Now, we are going to solve the Dirac equation for a neutral particle
with
anomalous electric interaction. For an electric field along the z axis
we
have that the Dirac equation takes the form
\begin{equation}
\label{electr}\left( \gamma ^0\partial _t+\gamma ^1\partial _r+{\frac
\gamma
r}^2\partial _\vartheta +\gamma ^3\partial _z+m+g\gamma ^3\gamma
^0E\right)
\Psi ~=0
\end{equation}
Following the pairwise scheme of separation, we have that Eq.
(\ref{electr})
can be reduced to a sum of two commuting first order differential
operators

\begin{equation}
\label{unoE}\left( \gamma ^0\gamma ^1\gamma ^3\partial _r\ +\frac{\gamma
^0\gamma ^2\gamma ^3}r\partial _\vartheta \ +gE\ -\lambda \right) \Phi
=0
\end{equation}
\begin{equation}
\label{dosE}\left( \gamma ^3\partial _0+\gamma ^0\partial _z+m\gamma
^3\gamma ^0+\lambda \right) \Phi =0
\end{equation}
with $\Phi =\gamma ^3\gamma ^0\Psi $ .Choosing to work in the Dirac
matrices
representation given by the expression  (\ref{repr}) we have that
Eq.(\ref
{unoE}) can be written as:
\begin{equation}
\label{prima}\left( \sigma ^2\partial _r\ -\ \frac{\sigma ^1}r\partial
_\vartheta \ -(gE-\lambda )\right) \Theta =0
\end{equation}
\begin{equation}
\left( \sigma ^2\partial _r\ -\ \frac{\sigma ^1}r\partial _\vartheta \
+\
(gE-\lambda )\right) \chi =0
\end{equation}
with

\begin{equation}
\Phi =\left(
\begin{array}{c}
\Theta \\
\chi
\end{array}
\right) =\left(
\begin{array}{c}
\Theta \\
\frac{i(m+E)}{\lambda -k_z}\sigma ^3\Theta
\end{array}
\right) ,\ \Theta =\left(
\begin{array}{c}
\Theta _1 \\
\Theta _2
\end{array}
\right)
\end{equation}
where the constant of separation $\lambda $ satisfies the relation:
$\lambda
^2=k_z^2+m^2-E^2.$ From (\ref{prima}) we have that the components of
$\Theta
$ satisfy the relation:
\begin{equation}
\left( \partial _r-\frac{k_\vartheta }r\right) \Theta _1=i(\lambda
-gE)\Theta _2
\end{equation}

\begin{equation}
\left( \partial _r+\frac{k_\vartheta }r\right) \Theta _2=-i(\lambda
-gE)\Theta _1
\end{equation}
the solution of this system reads
\begin{equation}
\Theta _1=cr^{\frac 12}Z_{k_\vartheta -1/2}(i(\lambda -gE)r)
\end{equation}
\begin{equation}
\Theta _2=-cr^{\frac 12}Z_{k_\vartheta +1/2}(i(\lambda -gE)r)
\end{equation}

\section{A solvable example}

The following example could be regarded as surprising because of the big
amount of functions appearing in the Dirac equation. This example seems
to
be contradictory since, following the remarks made at the introduction,
the
presence of additional matrices in the Dirac equation dramatically
reduces
the possibilities of separation of variables. But such a contradiction
does
not exist in the case to be presented. All the functions considered in
the
example, depend only on the radial variable $r$ namely in the form
$1/r$,
therefore their structure repeat the Lam\'e function arising when we use
spherical
coordinates. This one explains the restriction in the election of the
functions allowing complete separation of variables.

Then, let us consider a Dirac particle in a Coulomb field with anomalous
electric dipole interaction and also scalar and pseudoscalar
interactions.
At a first glance this problem seems artificial, because we do not
relate
the Coulomb term with the dipole one, but the problem includes the
possibility of a unified approach of a series of physical situations
like
the Hydrogen atom (Coulomb field), the non minimal electric dipole
interaction, and confinement model (scalar potential) and the quarkonium
theory\cite{Ravndal,Dominguez1,Dominguez2,Dominguez3}

The Dirac equation expressed in the diagonal tetrad gauge takes the form

\begin{equation}
\label{primo}\left\{ \gamma ^1\partial _r\ +\ \frac 1r(\gamma ^2\partial
_\vartheta +\frac{\gamma ^3}{\sin \vartheta }\partial _\varphi )+\gamma
^0(\partial _t-\frac{Ze^2}r)+q\gamma ^1\gamma ^0\frac \epsilon r+\zeta
\frac
Sr+\xi \gamma ^5\frac{P(\vartheta ,\varphi )}r\right\} \Psi =0
\end{equation}
here $Ze^2/r$ is the Coulomb potential, $q\epsilon /r$ is the electric
dipole interaction, $S$ is the scalar potential, $P$ is the pseudoscalar
potential, and $\zeta $ and $\xi $ are constants. After separating the
angular dependence, (\ref{primo}) takes the form
\begin{equation}
\label{second}\left\{ \gamma ^0\partial _r+\frac{k-q\epsilon }r-\gamma
^1(\partial _t-\frac{Ze^2}r)+\gamma ^1\gamma ^0m+\zeta \gamma ^1\gamma
^0\frac Sr\right\} \Phi =0
\end{equation}
\begin{equation}
\label{third}\left( \gamma ^2\gamma ^1\gamma ^0\partial _\vartheta
+\frac{%
\gamma ^3\gamma ^1\gamma ^0}{\sin \vartheta }\partial _\varphi -\gamma
^2\gamma ^3\xi P(\vartheta ,\varphi )\right) \Phi =k\Phi
\end{equation}

\noindent where $\Psi =\gamma ^1\gamma ^0\Phi $

We will not solve the equation (\ref{third}) governing the angular
dependence of the problem because this one is not possible without
knowing
the $P(\vartheta ,\varphi )$ function. Obviously, the solution for the
particular case $P=0$ will be expressed in terms of spherical harmonics
like
the Coulomb case.

Choosing to work in the representation
\begin{equation}
\label{repdir}\gamma ^0~=~\pmatrix{i\sigma^{3}& 0\cr 0&
i\sigma^{3}\cr},\
\gamma ^1~=~\pmatrix{\sigma ^{1}& 0\cr 0& -\sigma^{1}\cr},\ \gamma
^2~=~%
\pmatrix{\sigma ^{2}& 0\cr 0& -\sigma^{2}\cr},\ \ \gamma ^3~=~%
\pmatrix{0&1\cr 1&0 \cr}
\end{equation}

\noindent equation (\ref{second}) takes the form,

\begin{equation}
\label{fourth}\left[ i\sigma ^3\partial _r+\frac{k-\epsilon q}r+\sigma
^1(\partial _t-i\frac{Ze^2}r)+\sigma ^2(\zeta \frac Sr+m)\right] \Phi
_1=0
\end{equation}
substituting (\ref{repdir}) into (\ref{fourth}) we arrive at,

\begin{equation}
\label{pier}\left\{ \frac d{dr}-i\frac{k-\epsilon q}r\right\} \Theta
_1+\left( -(E+m)-\frac 1r(Ze^2+\zeta S)\right) \Theta _2=0
\end{equation}

\begin{equation}
\label{vt}\left\{ \frac d{dr}+i\frac{k-\epsilon q}r\right\} \Theta
_2+\left(
(E-m)+\frac 1r(Ze^2-\zeta S)\right) \Theta _1=0
\end{equation}
where $\Theta _1$ and $\Theta _2$ are the components of the spinor $\Phi
_1$%
\begin{equation}
\Phi _1=\left(
\begin{array}{c}
\Theta _1 \\
\Theta _2
\end{array}
\right)
\end{equation}
Introducing the notation $i(k-\epsilon q)=M,$ $m+E=A,$ $m-E=B,$ $\zeta
S+Ze^2=\alpha ,$ $Ze^2-\zeta S=\beta $, and $rD=\rho $ we have that eq.
(\ref
{pier}), (\ref{vt}) take the form
\begin{equation}
\label{tr}\left\{ \frac d{d\rho }-\frac M\rho \right\} \Theta _1+\left(
-\frac AD-\frac \alpha \rho \right) \Theta _2=0
\end{equation}
\begin{equation}
\label{che}\left\{ \frac d{d\rho }+\frac M\rho \right\} \Theta _2+\left(
-\frac BD+\frac \beta \rho \right) \Theta _1=0
\end{equation}

\noindent  we shall look for solutions of the system (\ref{tr}),
(\ref{che})
in the form of power series:

\begin{equation}
\label{serie}\Theta _1=e^{-\rho }\sum_{\upsilon =0}^\infty \rho
^{s+\upsilon
}a_\upsilon ,\quad \Theta _2=e^{-\rho }\sum_{\upsilon =0}^\infty \rho
^{s+\upsilon }b_\upsilon
\end{equation}
Substituting (\ref{serie}) into (\ref{tr}), (\ref{che}) and putting the
coefficients of $\rho ^{s+\upsilon +1}$ equal to zero we find
\begin{equation}
\label{un}(s+M)a_0+\beta b_0=0
\end{equation}
\begin{equation}
\label{deux}\alpha a_0-(s-M)b_0=0
\end{equation}
\begin{equation}
\label{trois}-a_{\upsilon -1}+(s+\upsilon +M)a_\upsilon -\frac
BDb_{\upsilon
-1}+\beta b_\upsilon =0
\end{equation}
\begin{equation}
\label{quatre}b_{\upsilon -1}-\ (s+\upsilon -M)b_\upsilon +\frac
ADa_{\upsilon -1}+\alpha b_\upsilon =0
\end{equation}
from (\ref{un}) and (\ref{deux}) it follows that the parameter $s$ takes
the
value
\begin{equation}
s=(M^2-\alpha \beta )^{1/2}
\end{equation}
where we have considered a positive root in order to avoid a divergent
solution at the origin. From (\ref{trois}) and (\ref{quatre}) we find
the
following relation between the coefficients $a_\upsilon $ and
$b_\upsilon $
\begin{equation}
\label{rela}\left\{ D(s+\upsilon +M)+B\alpha \right\} a_\nu =\left\{
B(s+\upsilon -M)-D\beta \right\} b_\upsilon
\end{equation}
The series (\ref{serie}) will have a good behavior at infinity if they
terminate for a finite value $N$ Putting $a_{N+1}=b_{N+1}=0$ in
(\ref{trois}%
) and (\ref{quatre}) with $a_N\neq 0$ and $b_N\neq 0$ we arrive at
\begin{equation}
\label{coci}\frac{a_N}{b_N}=-\frac BD
\end{equation}
then, from (\ref{rela}), (\ref{coci}) and the definitions of $B$ and $D$
we
arrive at:
\begin{equation}
\label{spectr}EZe^2=(s+N)\sqrt{m^2-E^2}+m\zeta S
\end{equation}
The expression (\ref{spectr}) is the condition of quantization of the
energy, and its structure is very similar to the one obtained for the
hydrogen atom. From (\ref{spectr}) can obtain some particular cases: a)
$%
\epsilon =0,$ $S=0$ Hydrogen atom. b) $\epsilon =0,$ $Z=0$ Confinement
of
quarks by a scalar potential c) $Z=0,$ $S=0.$ No minimal electric dipole
interaction.

Obviously, exact solutions of the Dirac equation containing such set of
terms like (\ref{primo}) is not only of physical interest, but also is
interesting from a mathematical point of view. Regretfully the amount of
examples of this kind are relatively scarce and therefore not exact
solutions are known from most of the problems.

\section{Concluding remarks}

In this article we have obtained some exact solutions of the Dirac
equation
for electric neutral Dirac particle with anomalous magnetic moment. The
results presented in this paper show the capabilities of the algebraic
method of separation of variables provided a suitable representation for
the
gamma matrices. The separation of variables for other curvilinear system
of
coordinates was not possible because of the matrix character of the
anomalous interaction in the Dirac equation. This peculiarity
dramatically
restricts the possibilities of finding exact solution in other geometric
and
physical configurations. Also, it would be of interest to find exact
solutions when other interactions are present.

\vspace{0.4cm}

\acknowledgments

\noindent One of the authors (G.V.S) wishes to express his gratitude to
the
Fund for Fundamental Investigations of the Republic of Byelorrussia for
financial support.

\end{document}